\documentclass[RNAAS]{aastex631}
\usepackage{amsmath}
\usepackage{hyperref}

\begin{document}

\title{A Spectroscopic Hunt for Post-Red Supergiants in the Large Magellanic Cloud I: Preliminary Results}

\author[0009-0007-4983-9850]{Kaitlyn M. Chen}
\affiliation{Harvey Mudd College \\
340 E Foothill Blvd. \\
Claremont, California 91711, USA}
\affiliation{Observatories of the Carnegie Institution for Science \\
813 Santa Barbara Street \\
Pasadena, CA 91101, USA}

\correspondingauthor{Kaitlyn M. Chen}
\email{kaichen@g.hmc.edu}

\author[0000-0003-3601-3180]{Trevor Z. Dorn-Wallenstein}\thanks{Carnegie Fellow}
\affiliation{Observatories of the Carnegie Institution for Science \\
813 Santa Barbara Street \\
Pasadena, CA 91101, USA}

\begin{abstract}

Yellow supergiants (YSGs) are rare and poorly understood, and studying them is critical to constraining massive star evolution. We obtained flux-calibrated Magellan Inamori Kyocera Echelle (MIKE) high-resolution spectra of 40 YSGs in the Large Magellanic Cloud (LMC); this sample likely contains post-red supergiants (RSGs). Fitting these data with ATLAS9 model atmospheres, we determined fundamental parameters for these stars. We measure the first spectroscopic luminosities for YSGs above 20 $M_\odot$, providing us a novel probe of the luminosity-to-mass ratio. Many stars in our sample appear to have anomalously high surface gravities, despite being confirmed LMC supergiants. We manually inspected our data finding evidence for binary companions and ongoing mass loss. Our work demonstrates the valuable role of high-resolution spectroscopy in interpreting the evolutionary status of cool supergiants.
\end{abstract}

\section{Introduction} \label{sec:intro}
Stars that begin their lives between $\sim$8 and 30 $M_\odot$ are expected to explode as hydrogen-rich supernovae from RSG progenitors \citep{2012A&A...537A.146E}. However, directly-imaged progenitors only span the range $\lesssim20$ $M_\odot$. This statistically robust discrepancy is known as the RSG problem \citep{2015PASA...32...16S, 2022MNRAS.515..897R}. The missing SNe between 20 and 30 $M_\odot$ have called the currently understood paradigm of massive star evolution into question. One proposed solution to the RSG problem is that before core-collapse, luminous RSGs lose a large amount of mass causing them to evolve bluewards and become post-RSGs. Validating this solution requires identifying post-RSGs, a challenging task that can be accomplished with high-resolution spectroscopy. 

\cite{DW22} identified a new class of pulsators, fast yellow pulsating supergiants (FYPS), whose pulsations are linked to their proposed status as post-RSG objects. To test this hypothesis, we selected 40 YSGs from \cite{DW22} with photometric luminosities $\log L/L_\odot \gtrsim5.0$, corresponding to the missing progenitors' mass range. Fourteen of the observed stars belong to the proposed FYPS class. Even if FYPS are an artifact of TESS systematics \citep{Pederson2023}, $\sim30\%$ of luminous YSGs are expected to be post-RSGs \citep{2016ApJ...825...50G}, making this a rich sample to identify and characterize these objects. 

\section{Methods} \label{sec:style}
We observed our targets using the MIKE spectrograph at the Magellan telescopes at Las Campanas Observatory. The data range from 3500-9500 \r{A} and were taken at a resolution of $R\sim58000$/$R\sim45500$ (red/blue). Spectral orders were flux calibrated before being stitched together. Further details about these observations will be published in forthcoming work.

\par
We first manually inspected the high-resolution spectra looking for emission at H$\alpha$, indicating past or ongoing mass loss. 70\% of our sample showed such emission with complex P Cygni profiles.

\par
We then smoothed the data down to a linear resolution of 20 \r{A} using a box kernel, which matches the spectral resolution of the ATLAS9 Models \citep{2003IAUS..210P.A20C}. We fit the data, varying the effective temperature, surface gravity, radial velocity, reddening, and radius\footnote{which we can constrain with our flux-calibrated data, given the known distance to the LMC}, using {\sc stsynphot}. For reddening, we used the average interstellar extinction curve towards the LMC \citep{Gordon2003}.

\par
We set uniform priors: $T_{\rm eff}\sim\mathcal{U}(4000,12500)$, $\log g\sim\mathcal{U}(-.5,3)$, $E(B-V)\sim\mathcal{U}(0,2.5)$, $R/R_\odot \sim\mathcal{U}(1,1000)$. We set a Gaussian prior for our radial velocity: $\mathcal{N}({\mu_{\rm lit.}, {\textrm{ 30 km s}}}^{-1})$ where $\mu_{\rm lit.}$ is the literature radial velocity for each star.

We ran Markov Chain Monte Carlo (MCMC) simulations using {\sc emcee} \citep{Foreman-Mackey2013} to sample from the posterior probability function, using the log-likelihood function: 
\begin{equation}
    \ln \mathcal{L} = -\frac{1}{2}\sum_i \frac{(f_i-\hat{f}_i)^2}{\sigma_i^2}
\end{equation}
with the model flux ($\hat{f}_i$) and the priors defined above. To estimate the observed flux errors ($\sigma_i$), we compute the standard deviation of the flux ($f_i$) in a five-pixel rolling window. We use 32 walkers, and run the sampler for 5000 steps. We manually inspected the results, discarding the burn-in (typically 1000 steps). We took the median sample for each parameter as the converged value.

\par

With these best-fit parameters, we plugged the effective temperatures and radii into the Stefan-Boltzmann law to find bolometric luminosities. We tabulate the results from the MCMC runs in Table \ref{table1}.

\par
We then inspected the fit residuals and original flux-calibrated data to determine the binarity of each star based on if there is an upturn in the blue, around $\sim$3000 to 4000 \r{A} \citep{Neugent2018}.

\par

\section{Results} \label{sec:floats}

70.0$\%$ (28/40) of stars in our sample exhibit emission at the H$\alpha$ line. The CSM indicated by this emission could be from mass loss, implying that the stars could have lost their envelope and evolved blueward. However, some stars like HD 33579 are confirmed to be pre-RSG and still exhibit strong mass loss because of their dynamically unstable atmosphere \citep{Nieuwenhuijzen2000, Humphreys2013}. All H$\alpha$ emitters show a complicated mix of absorption and emission at different velocities and widths, indicating an optically thick wind.

Notably, 92.8$\%$ (13/14) of the FYPS have an H$\alpha$ in emission. The difference between this and the whole sample is marginally statistically significant with a p-value of $\sim$0.0814, derived using Fisher's exact test \citep{Fisher92}. While H$\alpha$ is not a definitive indicator of a star's post-RSG status, our results demonstrate that strong mass loss plays a key role in the YSG phase. We stress the importance of properly treating YSG mass loss in massive star evolutionary models \citep{Glebbeek2009}.

Of the 40 stars, 20 show a clear blue excess indicative of an OB star companion, with a binary fraction of $50 \pm7.8\%$, consistent with previous findings in the LMC \citep{2013A&A...550A.107S}. We note that our reported value likely contains both false-positives and undetected companions due to the lower signal-to-noise in the blue. However, our results reaffirm the importance of binary stars in all phases of massive star evolution.

We can also measure the spectroscopic luminosity, $\mathcal{L} = T_{\rm eff}^4 /g$ \citep{2014A&A...564A..52L}, which directly probes the luminosity-to-mass ratio. We observe our sample to have consistently low spectroscopic luminosities in comparison to evolutionary tracks for stars of that mass range. This result could indicate that these stars are over-massive relative to their luminosities. We speculate that this could be due to a merger or mass gain in a previous binary interaction, though further work is necessary to validate this scenario and determine which of these stars are post-RSGs. In particular, because RSGs are fully convective, they dredge up CNO-processed material that can be detected in the post-RSG phase. We plan to conduct a full line-by-line analysis to measure surface abundances for carbon, nitrogen, and oxygen, which we will use to conclusively identify the post-RSGs in our sample.

\begin{deluxetable*}{lcccccccc}
\tabletypesize{\scriptsize}
\tablecaption{Atmospheric parameters and supplemental data for all 40 yellow supergiants in our sample}
\tablehead{\colhead{CommonName} & \colhead{FYPS} & \colhead{Binary} & \colhead{H$\alpha$ in emission} & \colhead{$\log T_{\rm eff}$} & \colhead{$\log g$} & \colhead{$R/R_\odot$} & \colhead{$\log L/L_\odot$} & \colhead{$\log \mathcal{L}/\mathcal{L}_\odot$}} 
\label{table1}
\startdata
HD 269879 & False & True & True & $3.704\substack{+0.003 \\ -0.003}$ & $0.01\substack{+0.02 \\ -0.01}$ & $387\substack{+6 \\ -6}$ & $4.948\substack{+0.004 \\ -0.003}$ & $4.197\substack{+0.010 \\ -0.021}$ \\ 
HD 269857 & False & False & True & $3.831\substack{+0.021 \\ -0.022}$ & $2.58\substack{+0.07 \\ -0.08}$ & $288\substack{+6 \\ -6}$ & $5.200\substack{+0.090 \\ -0.093}$ & $2.141\substack{+0.074 \\ -0.067}$ \\ 
HD 270050 & False & True & False & $3.809\substack{+0.005 \\ -0.004}$ & $2.00\substack{+0.03 \\ -0.02}$ & $212\substack{+4 \\ -4}$ & $4.846\substack{+0.007 \\ -0.005}$ & $2.625\substack{+0.021 \\ -0.028}$ \\ 
HD 269723 & True & False & True & $3.694\substack{+0.010 \\ -0.009}$ & $0.00\substack{+0.01 \\ -0.00}$ & $734\substack{+17 \\ -17}$ & $5.460\substack{+0.041 \\ -0.040}$ & $4.162\substack{+0.005 \\ -0.007}$ \\ 
HD 269662 & False & True & True & $3.835\substack{+0.002 \\ -0.002}$ & $3.00\substack{+0.00 \\ -0.00}$ & $169\substack{+2 \\ -2}$ & $4.751\substack{+0.003 \\ -0.003}$ & $1.733\substack{+0.002 \\ -0.002}$ \\ 
HD 269070 & False & False & True & $3.710\substack{+0.005 \\ -0.005}$ & $0.65\substack{+0.06 \\ -0.06}$ & $367\substack{+10\\ -10}$ & $4.922\substack{+0.007 \\ -0.007}$ & $3.582\substack{+0.063 \\ -0.063}$ \\ 
SK -69   99 & True & True & True & $3.888\substack{+0.003 \\ -0.003}$ & $3.00\substack{+0.00 \\ -0.01}$ & $87\substack{+1 \\ -1}$ & $4.382\substack{+0.006 \\ -0.005}$ & $1.950\substack{+0.007 \\ -0.003}$ \\ 
HD 269697 & False & False & False & $3.797\substack{+0.005 \\ -0.004}$ & $1.53\substack{+0.06 \\ -0.05}$ & $327\substack{+5 \\ -5}$ & $5.172\substack{+0.018 \\ -0.010}$ & $3.046\substack{+0.051 \\ -0.056}$ \\ 
HD 269902 & False & False & True & $3.798\substack{+0.002 \\ -0.002}$ & $3.00\substack{+0.00 \\ -0.00}$ & $301\substack{+2 \\ -2}$ & $5.105\substack{+0.002 \\ -0.002}$ & $1.585\substack{+0.001 \\ -0.001}$ \\ 
$[$W60$]$ D17 & False & True & False & $3.603\substack{+0.002 \\ -0.002}$ & $0.00\substack{+0.00 \\ -0.00}$ & $646\substack{+13\\ -13}$ & $4.986\substack{+0.009 \\ -0.009}$ & $3.803\substack{+0.001 \\ -0.001}$ \\ 
HD 269331 & False & False & True & $3.823\substack{+0.006 \\ -0.006}$ & $2.58\substack{+0.06 \\ -0.06}$ & $267\substack{+6 \\ -6}$ & $5.099\substack{+0.009 \\ -0.008}$ & $2.101\substack{+0.056 \\ -0.056}$ \\ 
HD 269661 & True & True & True & $3.864\substack{+0.003 \\ -0.003}$ & $3.00\substack{+0.00 \\ -0.00}$ & $160\substack{+2 \\ -2}$ & $4.819\substack{+0.004 \\ -0.003}$ & $1.849\substack{+0.002 \\ -0.001}$ \\ 
HD 268727 & False & False & True & $3.885\substack{+0.004 \\ -0.004}$ & $2.99\substack{+0.00 \\ -0.01}$ & $103\substack{+2 \\ -2}$ & $4.517\substack{+0.006 \\ -0.006}$ & $1.938\substack{+0.009 \\ -0.004}$ \\ 
HD 269762 & True & False & False & $3.916\substack{+0.000 \\ -0.000}$ & $1.00\substack{+0.00 \\ -0.00}$ & $116\substack{+2 \\ -2}$ & $4.753\substack{+0.013 \\ -0.012}$ & $4.058\substack{+0.000 \\ -0.000}$ \\ 
HD 269953 & True & False & True & $3.736\substack{+0.004 \\ -0.004}$ & $0.16\substack{+0.06 \\ -0.06}$ & $465\substack{+9 \\ -9}$ & $5.233\substack{+0.006 \\ -0.006}$ & $4.178\substack{+0.058 \\ -0.059}$ \\ 
HD  33579 & True & False & True & $3.851\substack{+0.003 \\ -0.003}$ & $3.00\substack{+0.00 \\ -0.00}$ & $375\substack{+4 \\ -4}$ & $5.505\substack{+0.003 \\ -0.003}$ & $1.797\substack{+0.002 \\ -0.001}$ \\ 
SV* HV  2450 & False & True & False & $3.684\substack{+0.005 \\ -0.005}$ & $2.97\substack{+0.02 \\ -0.04}$ & $1000\substack{+2 \\ -1}$ & $5.692\substack{+0.018 \\ -0.018}$ & $1.155\substack{+0.041 \\ -0.019}$ \\ 
SV* HV   883 & False & True & True & $3.628\substack{+0.003 \\ -0.005}$ & $0.00\substack{+0.00 \\ -0.00}$ & $293\substack{+7 \\ -8}$ & $4.401\substack{+0.010 \\ -0.009}$ & $3.903\substack{+0.002 \\ -0.003}$ \\ 
HD 269604 & True & True & True & $3.867\substack{+0.003 \\ -0.003}$ & $3.00\substack{+0.00 \\ -0.01}$ & $149\substack{+2 \\ -2}$ & $4.769\substack{+0.004 \\ -0.004}$ & $1.863\substack{+0.006 \\ -0.003}$ \\ 
HD 268819 & False & False & True & $3.784\substack{+0.003 \\ -0.003}$ & $1.49\substack{+0.03 \\ -0.03}$ & $355\substack{+4 \\ -4}$ & $5.192\substack{+0.005 \\ -0.004}$ & $3.036\substack{+0.029 \\ -0.025}$ \\ 
2MASS J05344326-6704104 & False & True & False & $3.602\substack{+0.001 \\ -0.000}$ & $0.00\substack{+0.00 \\ -0.00}$ & $551\substack{+11\\ -14}$ & $4.846\substack{+0.023 \\ -0.017}$ & $3.800\substack{+0.001 \\ -0.001}$ \\ 
CD-69   310 & True & False & True & $3.837\substack{+0.027 \\ -0.018}$ & $2.36\substack{+0.23 \\ -0.06}$ & $192\substack{+11\\ -8}$ & $4.872\substack{+0.102 \\ -0.088}$ & $2.381\substack{+0.062 \\ -0.220}$ \\ 
SP77  31-16 & False & True & True & $3.602\substack{+0.000 \\ -0.000}$ & $0.00\substack{+0.00 \\ -0.00}$ & $923\substack{+28\\ -28}$ & $5.293\substack{+0.027 \\ -0.027}$ & $3.800\substack{+0.001 \\ -0.001}$ \\ 
CPD-69   496 & False & False & False & $3.749\substack{+0.003 \\ -0.003}$ & $0.33\substack{+0.06 \\ -0.06}$ & $261\substack{+4 \\ -4}$ & $4.786\substack{+0.006 \\ -0.006}$ & $4.062\substack{+0.063 \\ -0.064}$ \\ 
HD 269781 & True & False & True & $3.875\substack{+0.003 \\ -0.003}$ & $3.00\substack{+0.00 \\ -0.00}$ & $229\substack{+2 \\ -2}$ & $5.174\substack{+0.004 \\ -0.004}$ & $1.892\substack{+0.001 \\ -0.001}$ \\ 
HD 268687 & True & False & True & $3.794\substack{+0.013 \\ -0.016}$ & $2.99\substack{+0.01 \\ -0.02}$ & $261\substack{+5 \\ -5}$ & $4.965\substack{+0.058 \\ -0.068}$ & $1.583\substack{+0.019 \\ -0.010}$ \\ 
SP77  48-6 & False & True & False & $3.667\substack{+0.014 \\ -0.014}$ & $0.72\substack{+0.04 \\ -0.04}$ & $700\substack{+29\\ -28}$ & $5.312\substack{+0.034 \\ -0.034}$ & $3.345\substack{+0.039 \\ -0.039}$ \\ 
HD 268828 & False & True & False & $3.716\substack{+0.011 \\ -0.010}$ & $1.46\substack{+0.09 \\ -0.09}$ & $224\substack{+7 \\ -7}$ & $4.517\substack{+0.050 \\ -0.046}$ & $2.794\substack{+0.087 \\ -0.093}$ \\ 
HD 269651 & True & True & True & $3.841\substack{+0.003 \\ -0.003}$ & $3.00\substack{+0.00 \\ -0.01}$ & $147\substack{+2 \\ -2}$ & $4.655\substack{+0.005 \\ -0.004}$ & $1.761\substack{+0.007 \\ -0.004}$ \\ 
HD 269982 & False & False & True & $3.829\substack{+0.010 \\ -0.011}$ & $2.99\substack{+0.01 \\ -0.01}$ & $227\substack{+4 \\ -4}$ & $4.983\substack{+0.050 \\ -0.052}$ & $1.717\substack{+0.010 \\ -0.007}$ \\ 
SK -69  148 & False & True & True & $3.653\substack{+0.001 \\ -0.001}$ & $0.00\substack{+0.00 \\ -0.00}$ & $475\substack{+4 \\ -4}$ & $4.920\substack{+0.005 \\ -0.005}$ & $4.004\substack{+0.001 \\ -0.002}$ \\ 
HD 268949 & False & True & True & $4.070\substack{+0.001 \\ -0.001}$ & $2.00\substack{+0.00 \\ -0.00}$ & $57\substack{+1 \\ -1}$ & $4.745\substack{+0.016 \\ -0.015}$ & $3.671\substack{+0.001 \\ -0.002}$ \\ 
HD 268946 & True & False & True & $3.877\substack{+0.003 \\ -0.003}$ & $3.00\substack{+0.00 \\ -0.00}$ & $203\substack{+2 \\ -2}$ & $5.078\substack{+0.004 \\ -0.004}$ & $1.901\substack{+0.001 \\ -0.001}$ \\ 
HD 269840 & True & False & True & $3.798\substack{+0.019 \\ -0.016}$ & $2.00\substack{+0.07 \\ -0.06}$ & $327\substack{+8 \\ -8}$ & $5.176\substack{+0.090 \\ -0.082}$ & $2.582\substack{+0.057 \\ -0.062}$ \\ 
RM 1-77 & False & True & True & $3.613\substack{+0.005 \\ -0.004}$ & $0.01\substack{+0.01 \\ -0.00}$ & $431\substack{+11\\ -12}$ & $4.678\substack{+0.021 \\ -0.014}$ & $3.840\substack{+0.005 \\ -0.009}$ \\ 
HD 269110 & False & True & True & $3.709\substack{+0.005 \\ -0.005}$ & $0.77\substack{+0.07 \\ -0.07}$ & $375\substack{+12\\ -11}$ & $4.939\substack{+0.009 \\ -0.009}$ & $3.454\substack{+0.074 \\ -0.072}$ \\ 
HD 269787 & True & False & True & $3.877\substack{+0.003 \\ -0.003}$ & $3.00\substack{+0.00 \\ -0.00}$ & $166\substack{+2 \\ -2}$ & $4.903\substack{+0.004 \\ -0.004}$ & $1.904\substack{+0.002 \\ -0.002}$ \\ 
HD 269807 & False & True & False & $3.865\substack{+0.004 \\ -0.005}$ & $2.99\substack{+0.01 \\ -0.02}$ & $157\substack{+2 \\ -3}$ & $4.805\substack{+0.010 \\ -0.007}$ & $1.865\substack{+0.019 \\ -0.009}$ \\ 
HD 268971 & False & False & False & $4.070\substack{+0.000 \\ -0.001}$ & $2.00\substack{+0.00 \\ -0.00}$ & $79\substack{+1 \\ -1}$ & $5.030\substack{+0.010 \\ -0.011}$ & $3.672\substack{+0.000 \\ -0.001}$ \\ 
HD 268865 & False & True & False & $3.733\substack{+0.003 \\ -0.003}$ & $1.89\substack{+0.08 \\ -0.09}$ & $229\substack{+4 \\ -4}$ & $4.605\substack{+0.007 \\ -0.007}$ & $2.438\substack{+0.090 \\ -0.083}$ \\ 
\enddata
\tablecomments{We expect our errors for the effective temperature, radius, and luminosity are underestimated and thus multiplied the errors by 10.
\newline Columns (left to right): Common Name, fast yellow pulsating supergiant, binary status, H$\alpha$ in emission, log effective temperature, surface gravity, radius, log luminosity, log spectroscopic luminosity}
\end{deluxetable*}
\newpage
\bibliography{rnaas}{}
\bibliographystyle{aasjournal}

\end{document}